\documentclass[10pt,conference,anonymous]{IEEEtran}
\IEEEoverridecommandlockouts

\newcommand{\tool}{Archer\xspace}
\newcommand{\alive}{\textsc{Alive2}\xspace}
\newcommand{\opt}{\textsc{opt}\xspace}
\newcommand{\llubi}{\textsc{LLUBI}\xspace}

\usepackage[T1]{fontenc}
\usepackage[utf8]{inputenc}

\usepackage{graphicx} 
\usepackage{xspace}
\usepackage{minted}
\usepackage{subcaption}
\usepackage{amsthm}
\usepackage{amsmath}
\usepackage{xcolor}
\usepackage{csquotes}
\usepackage{mathtools}
\usepackage{textcomp}
\usepackage{enumitem}
\usepackage{multirow}
\usepackage[most]{tcolorbox}
\usepackage{pifont}
\usepackage{makecell}
\usepackage[ruled,vlined,linesnumbered]{algorithm2e}
\usepackage{url}
\usepackage{booktabs}

\usepackage{tabularx}
\usepackage{array}
\usepackage{lstlinebgrd}

\newcolumntype{L}[1]{>{\raggedright\arraybackslash}p{#1}}
\newcolumntype{Y}{>{\raggedright\arraybackslash}X}

\usepackage{listings}
\usepackage{lstlinebgrd}
\usepackage{float}
\usepackage[hidelinks]{hyperref}

\definecolor{codebg}{RGB}{248,248,248}
\definecolor{codeframe}{RGB}{220,220,220}
\definecolor{codecomment}{RGB}{106,153,85}
\definecolor{codekeyword}{RGB}{0,92,197}
\definecolor{codestring}{RGB}{163,21,21}
\definecolor{callbg}{HTML}{E2EFFF}

\lstdefinelanguage{LLVMIR}{
  morekeywords={
    define,declare,global,constant,private,internal,external,
    dso_local,unnamed_addr,align,attributes,ret,br,switch,invoke,
    resume,unreachable,call,phi,select,icmp,fcmp,load,store,
    alloca,getelementptr,extractelement,insertelement,shufflevector,
    extractvalue,insertvalue,bitcast,addrspacecast,trunc,zext,sext,
    fptrunc,fpext,fptoui,fptosi,uitofp,sitofp,ptrtoint,inttoptr,
    add,fadd,sub,fsub,mul,fmul,udiv,sdiv,fdiv,urem,srem,frem,
    shl,lshr,ashr,and,or,xor,fneg,
    eq,ne,ugt,uge,ult,ule,sgt,sge,slt,sle,
    true,false,zeroinitializer,poison,null,undef,
    nsw,nuw,exact,inbounds,noundef,immarg,
    nocallback,nofree,nosync,nounwind,speculatable,willreturn,
    memory,none
  },
  sensitive=true,
  morecomment=[l]{;},
  alsoletter={\%@.\-_<>},
  morestring=[b]",
}

\lstdefinestyle{irstyle}{
  language=LLVMIR,
  basicstyle=\ttfamily,
  keywordstyle=\color{codekeyword}\bfseries,
  commentstyle=\color{codecomment}\itshape,
  stringstyle=\color{codestring},
  rulecolor=\color{codeframe},
  xleftmargin=6pt,
  xrightmargin=6pt,
  breaklines=true,
  breakatwhitespace=true,
  showstringspaces=false,
  columns=fullflexible,
  keepspaces=true,
  captionpos=b,
  numbers=none
}

\lstdefinestyle{irstyle2}{
  language=LLVMIR,
  basicstyle=\ttfamily\footnotesize,
  keywordstyle=\color{codekeyword}\bfseries,
  commentstyle=\color{codecomment}\itshape,
  stringstyle=\color{codestring},
  rulecolor=\color{codeframe},
  xleftmargin=6pt,
  xrightmargin=6pt,
  breaklines=true,
  breakatwhitespace=true,
  showstringspaces=false,
  columns=fullflexible,
  keepspaces=true,
  captionpos=b,
  numbers=none
}

\newcounter{promptbox}

\usepackage{xspace}

\newcommand{\ie}{\hbox{\emph{i.e.}}\xspace}

\newcommand{\para}[1]{\smallskip\noindent\textbf{#1}\xspace}
\newcommand{\tmark}{{\scriptsize\ding{228}}}

\newcommand{\circledone}[1]{{\Large\ding{172}}}
\newcommand{\circledtwo}[1]{{\normalsize\ding{173}}}
\newcommand{\circledthree}[1]{{\normalsize\ding{174}}}
\newcommand{\circledfour}[1]{{\normalsize\ding{175}}}
\newcommand{\circledfive}[1]{{\normalsize\ding{176}}}

\AtBeginDocument{%
  }

\begin{document}

\title{\tool: Towards Agentic Review for Compiler Optimizations}

\author{\IEEEauthorblockN{Yunbo Ni}
\IEEEauthorblockA{\textit{The Chinese University of Hong Kong} \\
ybni@cse.cuhk.edu.hk}
\and
\IEEEauthorblockN{Shaohua Li}
\IEEEauthorblockA{\textit{The Chinese University of Hong Kong} \\
shaohuali@cuhk.edu.hk}
}

\maketitle

\begin{abstract}
Modern compilers are frequently updated, but expert review capacity is highly limited, leading to delayed integration and, in some cases, subtle semantic bugs entering the compiler codebase. 
Automating the code review process with modern general code review agents may be feasible, but it faces critical challenges due to compiler complexity.
In this paper, we use LLVM as our target compiler and present \tool, the first automated agentic code review tool for compiler optimizations.
\tool constrains the agentic review process from both ends by using obligations to guide analysis and a deterministic validation guard to admit only findings backed by executable evidence.

We evaluated \tool on 70 open PRs and 328 closed PRs in LLVM from the last two months.
The review results are shocking and concerning: \tool discovers that \emph{21\%} of open PRs and \emph{11\%} of closed PRs are buggy, \ie, introducing semantic bugs such as miscompilations in LLVM.
Our findings expose a critical gap in the capacity for critical review in large compiler projects and demonstrate the practical value of \tool as an additional reviewer. 
\end{abstract}

\section{Introduction}
\label{sec:intro}
Compilers are fundamental to the modern software ecosystem and are typically developed by large, distributed communities. 
For example, LLVM, one of the most widely used and mature compiler infrastructures, has more than 5,000 contributors working on components, with hundreds of commits integrated into the codebase every day~\cite{online:llvm-contributors}. 
However, this vibrant contribution flow hides a subtle bottleneck in its review capacity. The volume and complexity of incoming patches can outpace maintainers' review bandwidth, leading to long review queues, delayed integration, and, in some cases, shallow or inconsistent feedback~\cite{online:llvm-status}.
As noted by a lead maintainer of the LLVM project~\cite{online:llvm-bad-parts}:
\emph{``lack of review capacity makes for a bad contributor experience, and can also result in bad changes making their way into the codebase.''}

This bottleneck is particularly problematic for compiler optimization patches.
Unlike many ordinary software changes, an optimization patch must preserve the semantics of the source program while interacting with a long sequence of analysis and canonicalizations.
Such missed semantic issues can lead to serious regression bugs~\cite{zhu2021regression}, particularly miscompilation bugs, which are among the most critical and hardest-to-detect defects in compilers~\cite{chen2021compilersurvey}. 
Even worse, experienced reviewers may overlook corner cases or fail to fully anticipate the impact of a change on downstream optimizations in a limited time. 

This naturally raises a research question: \textbf{\textit{can this review process be automated to reduce the burden on maintainers?}}
Compiler optimization code review differs fundamentally from traditional compiler testing: it is patch-specific, semantics-oriented, and must provide useful feedback within the short turnaround of a pull request. 
By contrast, compiler testing typically relies on longer-running exploration and largely unguided search~\cite{zhu2022fuzzingsurvey}.

Recent progress in code agents suggests the possibility of assisting code review by automatically commenting on a patch~\cite{cihan2025reviewsurvey}.
Emerging tools, including Codex~\cite{online:codex}, Github Copliot~\cite{online:copilot} and the specialized review tool CodeRabbit~\cite{online:coderabbit}, have begun to demonstrate this potential, mainly targeting general-purpose code review concerns such as style issues, maintainability problems, and common code smells. 
However, code review for compiler optimizations is substantially more demanding. 
Existing agentic review tools face two major challenges in this setting: (1) the gap between static textual context and complex semantics of compiler optimizations, and (2) the lack of executable and deterministic validation for textual suspicions. 
Below we elaborate on the two challenges:

\para{\tmark~Challenge 1: Gap between static textual context and complex semantics of compiler optimizations.}
Existing agentic review tools largely follow a text-centric formulation, where they inspect the patch and retrieve surrounding repository context to generate review comments~\cite{zhang2025laura}.
This formulation is useful for many implementation-level issues, but it is insufficient for compiler optimization review. 
A compiler optimization patch is part of a multi-stage semantic pipeline.
To review a single optimization patch, the reviewer must reason not only about the local rewrite, but also about the possible semantic conditions under which the rewrite remains valid. 
For example, in LLVM, even a simple arithmetic transformation in InstCombine pass may become incorrect only under a particular combination of poison propagation, signedness constraints or overflow flags~\cite{lopes2021alive2}.
These dependencies are semantic rather than syntactic.
They are therefore difficult to infer from textual exploration alone, even with retrieved repository context or API-level code navigation, because the relevant relation is not a call edge or a local data dependency, but a complex obligation at compiler semantic level.

\para{\tmark~Challenge 2: Lack of executable and deterministic validation for textual suspicions.}
Existing LLM-based review tools usually produce natural-language comments about a patch~\cite{sun2025study3}. 
For compiler optimization review, such textual comments are often insufficient. 
A verbose comment may point to a plausible bug, but without executable evidence, developers still need to determine whether the report is a real bug or a false positive. 
This burden is further amplified by the fact that LLMs can produce plausible but unfaithful explanations, rationalizing an incorrect conclusion with seemingly coherent analysis~\cite{turpin2023unfailthful,madsen2024self}.
This need for evidence is already reflected in real compiler review practice~\cite{online:instcombineguide, online:llvm-ai}, where LLVM developers are encouraged to accompany patches and comments with \alive and \opt evidence. 
However, constructing such artifacts is non-trivial.
A useful evidence must execute the specific optimization pass and further cover the changed lines of code in this certain patch, otherwise the result still leaves developers with additional manual validation work.
This makes our goal fundamentally different from ordinary compiler testing, which may accept any newly discovered bug.

\para{\tmark~Our design.}
The key idea is to position the agentic review process between compiler-specific semantic guidance on the input side and executable validation on the output side. 
This idea leads to two core mechanisms.
On the input side, modern code agents can already search large repositories, but compiler optimization review needs guidance about which semantic relations are worth checking.
Historical correctness fixes provide such guidance because they encode developers' experience about fragile semantic interactions.
Based on this observation, we propose \textit{dynamic obligation construction}, which distills historical fixes into semantic obligations that guide the agent's analysis beyond textual proximity or static context retrieval.
On the output side, the agent's analysis is still expressed in verbose natural language and may mix useful insights with incorrect reasoning.
We therefore design a \textit{deterministic validation guard} that converts textual analysis into executable, patch-specific evidence.
The guard requires the agent to formulate actionable validation strategies, use them to guide proof-of-concept generation, and report a finding only when the evidence exercises the relevant changed behavior in the patch.

Based on this design, we propose \tool, the first automated agentic code review tool for compiler optimizations, designed to review semantically sensitive compiler patches. 
We implement \tool on top of mini-SWE-agent~\cite{yang2024swe} by connecting its tool-call interface with LLVM-specific obligations and the deterministic validation guard.
The design is framework-agnostic and can be instantiated in other agent systems.

To evaluate the effectiveness of \tool, we study real-world compiler review scenarios in LLVM, one of the most mature open-source compiler infrastructures. 
We collected a total of 398 PRs (70 open and 328 closed) from the LLVM GitHub repository over the two months prior to the time of writing. \tool identifies 51 PRs that contain semantic bugs, meaning that in the past two months, \textbf{more than 21\% of open PRs and around 11\% of closed PRs in LLVM are buggy}.
This concerning finding points to a concrete review capacity gap, showing that semantic bugs are not only present in open PRs awaiting review, but also persist in closed PRs that have already undergone substantial manual review. 
This suggests that the challenge is not merely insufficient review volume, but the difficulty of sustaining semantics-aware review at the scale of a large compiler project. 
In conclusion, this paper makes the following contributions:

\begin{itemize}[labelwidth=!, labelindent=0pt, itemsep=5pt, topsep=2pt]

    \item We present \tool, the first agentic code review framework for compiler optimization patches, targeting semantic correctness bugs beyond generic review comments.
    
    \item We propose \textit{dynamic obligation construction} as input-side semantic guidance, distilling historical correctness fixes into obligations that direct the agent toward fragile compiler-optimization semantics.
    
    \item We design a \textit{deterministic validation guard} as output-side evidence control, allowing the agent to report a finding only when its textual analysis is converted into executable, patch-specific evidence.
    
    \item We implement and evaluate \tool on real LLVM PRs. Among 70 open and 328 closed PRs over the past two months, \tool finds 51 semantic bugs, showing its value for both ongoing review and bug discovery.
    
\end{itemize}

\tool has been open-source in \url{https://github.com/cuhk-s3/Archer}.
We believe that this work highlights a promising direction for integrating agents into the review workflow for large-scale infrastructure software. 

\begin{figure}[tp]
    \centering
    \includegraphics[width=\linewidth]{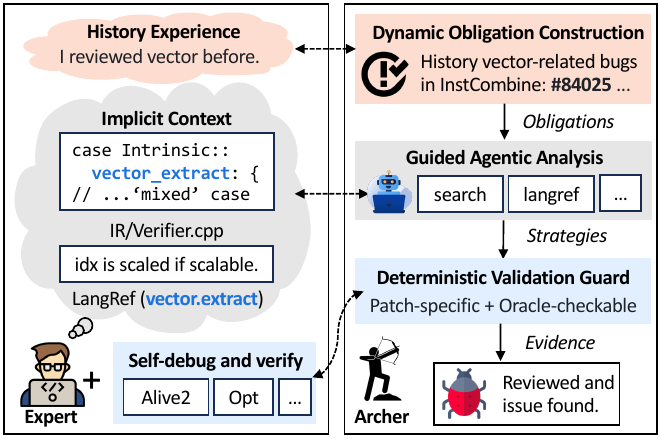}
    \vspace{-10pt}
    \caption{Example of how \tool conducts automated review on LLVM optimization PR in expert-like way.}
    \vspace{-10pt}
    \label{fig:motivation}
\end{figure}

\section{Illustrative Example and Motivation}
\label{sec:motivation}
We use a real InstCombine pull request\footnote{\url{https://github.com/llvm/llvm-project/pull/183329}} from LLVM to illustrate the kind of subtle semantic bugs targeted by \tool shown in Figure~\ref{fig:motivation}. 
This PR introduced a miscompilation bug in the LLVM, which was found by \tool. 
The core of this bug lies in the return-type semantics of the LLVM intrinsic \texttt{vector.extract}. 
In the following, we first describe how a human expert would review this PR and how this process motivates the design of \tool. 

\para{\tmark How human experts review this PR.} 
A human reviewer typically begins with prior development experience, accumulated from working on the relevant compiler components over time. 
In this example, a reviewer familiar with the InstCombine pass and past vector-related issues would likely pay particular attention to vector lane semantics, since many historical bugs have arisen from subtle mistakes in this area. 
This corresponds to the first part of the human expert thinking model shown in the middle of Figure~\ref{fig:motivation}.
Guided by this heuristic, the reviewer synthesizes \textit{an implicit context} that extends beyond the local diff. 
To build a comprehensive model of \texttt{vector.extract} element, they may cross-reference formal specifications from the LangRef 
with semantically coupled subsystems, such as the return-type edge cases in \texttt{IR/Verifier.cpp}. 
Finally, the reviewer actively verifies it using external tools like \alive and \opt, mirroring the actionable validation phases in Figure~\ref{fig:motivation}.

\para{\tmark How our \tool succeeds.}
The review workflow of \tool follows the same high-level structure, but makes the process explicit and reproducible, as illustrated in the right part of Figure~\ref{fig:motivation}.
Before review, \tool constructs pass-level obligations from historical correctness fixes through \textit{dynamic obligation construction}.
For this PR, the pre-built InstCombine obligations guide \tool to inspect whether the patch triggers vector lane semantics, rather than treating the change as a local implementation detail.
\tool then starts to review the PR and analyze this patch.
It calls on the search tools to collect patch-centered context like human experts and identifies several possible correctness issues. 
Under the \textit{deterministic validation guard}, \tool cannot report all the suspicions directly.
It first turns the analysis into actionable mutation strategies, including a strategy that varies the return type of \texttt{vector.extract} to challenge the transformation.
The guard then uses patch-related tests as seeds and guides \tool to mutate them into validation cases that still exercise the changed optimization.
To activate the suspected return-type corner case, \tool further instantiates concrete inputs and validates the resulting source-target behavior with compiler-aware oracles.
The final evidence shows that the patched compiler miscompiles a valid LLVM IR program.
\tool reports the bug with a reproducer and supporting semantic analysis, and the developer later confirmed the issue and fixed it within one day.

\begin{figure}
    \centering
    \includegraphics[width=\linewidth]{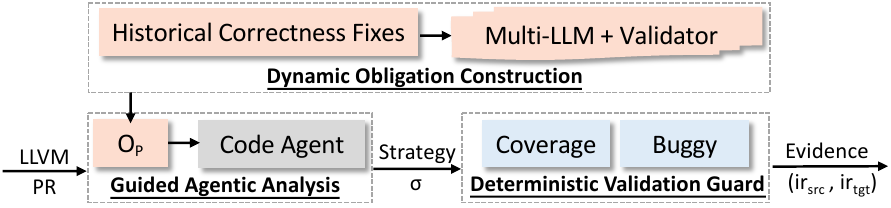}
    \caption{Design of high-level workflow for \tool.}
    \label{fig:design}
\end{figure}

\section{Design}
\label{sec:design}

Figure~\ref{fig:design} shows the high-level design of \tool. 
\tool's design constrains agentic compiler review with input-side semantic guidance and output-side executable validation, realized through \textit{dynamic obligation construction} and a \textit{deterministic validation guard}.
This section presents the full review process by first describing how obligations are constructed in Section~\ref{sec:design-obligation}, then how they guide agentic analysis in Section~\ref{sec:design-agent}, and finally how the guard turns analysis results into validated evidence in Section~\ref{sec:design-guard}.

\begin{algorithm}[tp]
\caption{Dynamic Obligation Construction}
\label{alg:pass-knowledge}
\SetFuncSty{textsc}
\DontPrintSemicolon
\SetKwInput{KwInput}{Input}
\SetKwInput{KwOutput}{Output}
\SetKwFunction{GroupByPass}{GroupByPass}
\SetKwFunction{Extract}{ExtractObligation}
\SetKwFunction{Generate}{GenerateIRPair}
\SetKwFunction{Verify}{ProofCheck}
\SetKwFunction{Summarize}{SummarizeObligations}
\SetKwProg{Proc}{procedure}{:}{}

\Proc{\textsc{BuildObligation}($D$)}{
    \KwInput{Historical correctness fixes $D$}
    \KwOutput{Pass-level obligation base $O$}

    $G \gets \GroupByPass(D)$; \textcolor{blue}{\tcp{fixes by pass}}
    $O \gets \emptyset$\;

    \ForEach{pass bucket $P \in G$}{
        $V_P \gets \emptyset$; \textcolor{blue}{\tcp{validated cases for $P$}}
        \ForEach{instance $x \in P$}{
            \textcolor{blue}{\tcp{Abstract: bug $\rightarrow$ obligation}}
            $o \gets \Extract(x.\mathit{issue}, x.\mathit{patch})$\;
            \textcolor{blue}{\tcp{Generate: obligation $\rightarrow$ IR}}
            $(ir_{src}, ir_{tgt}) \gets \Generate(o)$\;
            \tcp{\textcolor{blue}{Validate: verify recovery}}
            \If{\Verify($ir_{src}, ir_{tgt}$)}{
                append $(o, ir_{src}, ir_{tgt})$ to $V_P$; 
            }
        }
        \textcolor{blue}{\tcp{Summarize: high-level obligations by pass}}
        $O_P \gets \Summarize(V_P)$;
    }
    \Return{$O$}\;
}
\end{algorithm}

\subsection{Dynamic Obligation Construction}
\label{sec:design-obligation}

The goal of this component is to build an obligation base from historical correctness fixes that can guide future reviews with reusable compiler-semantic concerns.
An obligation describes a semantic relation that an optimization pass should preserve, rather than specific implementation details.
It consists of a set of semantic elements and transformation patterns that explains how these elements interact during optimization. 

\para{\tmark Recovering obligations with dynamic feedback.}
Existing general code review systems~\cite{yujia2025study1,sson2025study2,sun2025study3} often augment review generation with retrieved textual artifacts through Retrieval-Augmented Generation~\cite{lewis2020rag} (RAG).
Even when historical fixes are retrieved, they are typically used as static textual context. 
This is insufficient for compiler optimization review, as will be shown in our evaluation in Section~\ref{sec:eval-rq3}. The main reason is that it fails to recover the hidden semantics behind the implementation details. 
We take inspiration from compiler development practice to find the connection.
When developers fix an optimization bug, they often attach a reproducer to make the hidden semantic relation observable, which suggests a natural way to recover semantics.
Rather than treating a fix as a static patch, \tool uses reproducer generation as dynamic feedback to infer what semantic condition the fix was meant to preserve.
A candidate obligation is retained only if it can explain a real reproducer whose transformation exposes the same kind of semantic issue behind the historical fix.

\para{\tmark Organizing obligations by pass.}
We organize obligations at the granularity of optimization passes.
This organization follows the modular structure of modern compilers, where each pass implements a relatively specific class of transformations and maintains its own semantic assumptions.
It also matches real-world review responsibilities, since a compiler optimization PR is usually reviewed by experts of a certain pass.
Formally, \tool constructs an obligation base $O$, where each optimization pass $P$ is associated with a validated obligation set $O_P$.
During review, \tool identifies the affected pass and loads the corresponding $O_P$ as semantic guidance.
\begin{figure}[tp]
    \centering
    \includegraphics[width=0.98\linewidth]{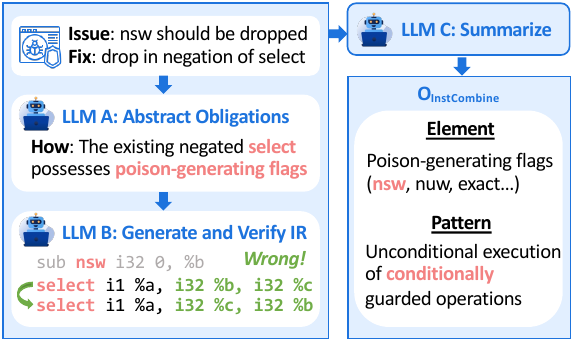}
    \caption{Example of how \tool automatically constructs pass-level obligations for InstCombine.}
    \vspace{-10pt}
    \label{fig:knowledge-example}
\end{figure}
For the pass-level obligation set $O_P$ to serve as effective guidance during review, its obligations should satisfy three properties:

\begin{itemize}[labelwidth=!, labelindent=5pt, itemsep=3pt, topsep=2pt, leftmargin=15pt]
    \item \textbf{High-level.} Pass-level obligations should not be tied to individual cases. Concrete bugs are often too specific and rare, making it difficult to match them directly to new patches. Compiler optimizations also continuously evolve, so obligations grounded in exact historical implementations can quickly become outdated. Therefore, constructed obligations should abstract away from case-specific details and capture reusable IR-level semantic elements and recurring transformation patterns.

    \item \textbf{LLM-readable.} The constructed obligations must be directly usable by LLM agents. Raw bug reports and fixes often contain much irrelevant information, such as outdated IR and LLVM source code, which is difficult for LLMs to read and reason about reliably~\cite{zhang2025llm4ir}. Directly providing such raw materials can also introduce excessive low-level context, making them ineffective as review guidance.

    \item \textbf{Precise.} The obligations extracted from historical fixes should capture the root causes of the original correctness issues. Otherwise, they may guide the agent toward irrelevant mutations or even misleading semantic concerns when reviewing new PRs.
\end{itemize}

To satisfy the above requirements, Algorithm~\ref{alg:pass-knowledge} uses a multi-LLM pipeline.
Given a collection of historical correctness fixes $D$, the algorithm works as follows:

\begin{enumerate}[label={\textbf{Step~\arabic*.}}, labelwidth=!, labelindent=30pt, itemsep=5pt, topsep=2pt]

    \item \emph{Group Historical Fixes by Pass} (line 2):
    The algorithm groups historical correctness fix instances $D$ by optimization pass to obtain pass-specific buckets $G$.
    This ensures that each obligation set $O_P$ is constructed from fixes related to the same transformation context.
    
    \item \emph{Abstract Candidate Obligations} (lines 7--8):
    For each fix instance $x$ in a pass bucket $P$, LLM A analyzes the fix to extract a candidate obligation $o$.
    At this stage, $o$ is only a textual hypothesis about what semantic condition the historical fix was meant to preserve.
    
    \item \emph{Generate Executable Reproducer} (lines 9--10):
    LLM B attempts to materialize each candidate obligation $o$ into an executable IR reproducer pair $(ir_{\text{src}}, ir_{\text{tgt}})$.
    This step uses a reproducer to recover the hidden obligation.
    
    \item \emph{Validate Candidate Obligations} (lines 11--12):
    The generated reproducer is validated with a compiler-aware proof checker like \alive.
    If $ir_{\text{src}}$ and $ir_{\text{tgt}}$ expose a semantic difference, the candidate obligation is treated as dynamically grounded and the triple $(o, ir_{\text{src}}, ir_{\text{tgt}})$ is retained in $V_P$.
    Otherwise, $o$ is discarded because it cannot be connected to executable compiler behavior.
    
    \item \emph{Summarize Pass-Level Obligations} (line 14):
    Finally, the validated candidates in $V_P$ are summarized by LLM C into the pass-level obligation set $O_P$.
    This step abstracts away case-specific details from individual fixes while preserving reusable semantic elements and transformation patterns that can guide future reviews.

\end{enumerate}

\begin{figure}
    \centering
    \includegraphics[width=0.9\linewidth]{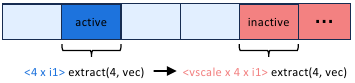}
    \caption{Example of semantic rationale in \tool's strategy.}
    \label{fig:strategy}
\end{figure}

At the end of the construction process, each optimization pass has a dedicated obligation base $O$ that captures reusable and semantically validated review guidance.

\para{\tmark Construction Example.}
Figure~\ref{fig:knowledge-example} illustrates Algorithm~\ref{alg:pass-knowledge} with an InstCombine example.
We use one instance to show how a historical bug is converted into a pass-level obligation.

For the selected instance $x \in P_{\texttt{InstCombine}}$, the underlying issue is a miscompilation caused by the interaction between a \texttt{select} rewrite and the poison-generating \texttt{nsw} flag.
LLM A abstracts the issue description and the fixing patch into a candidate obligation.
In this case, the obligation captures that a transformation involving a negated \texttt{select} must re-check whether poison-generating flags remain valid after the rewrite.
LLM B then instantiates this obligation into a candidate IR reproducer $(ir_{\text{src}}, ir_{\text{tgt}})$.
\alive confirms that the reproducer exposes a semantic mismatch, so the triple $(o, ir_{\text{src}}, ir_{\text{tgt}})$ is retained in $V_{\texttt{InstCombine}}$.
Finally, the validated set $V_{\texttt{InstCombine}}$ with other verified cases is summarized into $O_{\texttt{InstCombine}}$ by LLM C.
The resulting pass-level obligations capture reusable IR-level semantic elements and transformation patterns, including the role of \texttt{nsw} element and the fragility of \texttt{select}-based conditional rewrite pattern.
Overall, the example shows how dynamic obligation construction turns raw historical fixes into obligations that are high-level, LLM-readable, and precise.

\subsection{Guided Agentic Analysis}
\label{sec:design-agent}

Given a reviewed PR, \tool prepares a patch-specific review harness with the changed compiler source tree, the affected optimization pass, and its pass-level obligations $O_P$. 
The agent explores the compiler codebase through standard search and analysis actions, while using $O_P$ to focus on relevant semantic relations rather than open-ended repository context.
To support compiler-specific reasoning, \tool also provides access to the LLVM Language Reference~\cite{online:llvm-langref} and \opt, allowing the agent to check IR semantics and validate intermediate analysis during exploration.
The output of this stage is a set of actionable validation strategies.
We denote each strategy as $\sigma$, which specifies a mutation method, semantic rationale, and the expected validation result.
These strategies are then passed to the deterministic validation guard in Section~\ref{sec:design-guard}.

\para{\tmark Strategy Example.}
The following example illustrates an actionable strategy $\sigma$ produced by \tool.
Guided by VectorCombine obligations about vector scaling and lane, $\sigma$ contains:

\begin{itemize}[labelwidth=!, labelindent=5pt, itemsep=3pt, topsep=2pt, leftmargin=15pt]
\item \textbf{Mutation method.}
Make \texttt{vector.extract} extract a fixed vector from a scalable vector.

\item \textbf{Semantic rationale.}
As illustrated in Figure~\ref{fig:strategy}, in the concrete case of extracting \texttt{<4 x i1>} from \texttt{<vscale x 4 x i1>}, the patch may reason about the extracted lanes with an incorrect scaling assumption.
As a result, lanes that should remain active may be treated as inactive.

\item \textbf{Expected validation result.}
The optimization incorrectly replaces the extracted mask with a zero vector.
\end{itemize}

\begin{lstlisting}[style=irstyle2,float=tp,caption={Execution harness used by input-realized IR testing.},label={lst:difftest-template}]
define {ret_ty} @test_func({args}) {
    ...// generated IR 
}
define {ret_ty} @main(i32 %argc,ptr %argv) {
entry:
  %r = call {ret_ty} @test_func({args})
  ret {ret_ty} %r
}
\end{lstlisting}

\begin{algorithm}[tp]
\caption{Deterministic Validation Guard}
\label{alg:guard}
\SetFuncSty{textsc}
\DontPrintSemicolon
\SetAlgoNoLine
\SetKwInput{KwInput}{Input}
\SetKwInput{KwOutput}{Output}
\SetKwFunction{Materialize}{MaterializeCases}
\SetKwFunction{Transform}{Transform}
\SetKwFunction{ProofCheck}{ProofCheck}
\SetKwFunction{TestCheck}{TestCheck}
\SetKwFunction{PatchTriggered}{PatchTriggered}
\SetKwProg{Proc}{procedure}{:}{}

\Proc{\textsc{DeterministicGuard}($\sigma, T, C^{-}, C^{+}$)}{
\KwInput{Strategy $\sigma$, PR-related tests $T$, pre-patch compiler $C^{-}$, post-patch compiler $C^{+}$}
\KwOutput{Accepted validation case $c$ or $\bot$}

$Q \gets \Materialize(\sigma, T)$\;

\ForEach{source IR $ir_{src} \in Q$}{
    $ir_{tgt}^{+} \gets \Transform(C^{+}, ir_{src})$\; 

    $r \gets \ProofCheck(ir_{src}, ir_{tgt}^{+})$\;

    \If{$r$ is inconclusive}{
        $r \gets \TestCheck(ir_{src}, ir_{tgt}^{+}, \sigma)$\;
    }

    \If{$r$ exposes no semantic discrepancy}{
        \textbf{continue}\;
    }

    \If{\PatchTriggered($ir_{src}, C^{-}, C^{+}$)}{
        $c \gets (ir_{src}, ir_{tgt}^{+}, r)$\;
        \Return{$c$}\; 
    }
}

\Return{$\bot$}\;
}
\end{algorithm}

\subsection{Deterministic Validation Guard}
\label{sec:design-guard}

The goal of this component is to prevent the agent's textual analysis from being reported unless it can be reduced to deterministic compiler evidence ($ir_{src}$, $ir_{tgt}$) with properties:

\begin{itemize}[labelwidth=!, labelindent=5pt, itemsep=3pt, topsep=2pt, leftmargin=15pt]
    \item \textbf{Patch-triggering.} $ir_{src}$ must exercise the specific optimization pass and cover the changed behavior introduced by the PR. Otherwise, it may validate an unrelated transformation and cannot justify a finding for the current patch.
    
    \item \textbf{Oracle-checkable.} The source-target behavior between $ir_{src}$ and $ir_{tgt}$ must be checked by a compiler-aware oracle rather than by textual reasoning alone. 
\end{itemize}

\para{\tmark Ensuring patch triggering.}
Although an agent could synthesize LLVM IR tests from scratch, doing so is expensive and unreliable because generated IR often requires extensive repair before it can exercise the intended optimization.
Our design is based on a simpler observation: LLVM optimization-related PRs almost always come with tests~\cite{online:llvm-how-to-submit-a-patch}, and those tests already encode patch-relevant structure.
However, exercising the pass is still not enough.
A validation case should expose behavior introduced by the reviewed patch, rather than an unrelated latent bug in the same optimization pipeline.
To enforce this requirement, \tool compares the case under the unpatched compiler $C^{-}$ and the patched compiler $C^{+}$.
A case is considered patch-triggering only when the semantic discrepancy appears under $C^{+}$ but not under $C^{-}$.

\para{\tmark Complementary validation oracles.}
LLVM developers commonly use \alive to validate optimization correctness.
\tool follows this practice with \textsc{ProofCheck}, which checks source-target equivalence with \alive.
However, \alive is not complete for all real PRs, especially when patches involve unbounded loops, unsupported intrinsics, or solver-hard path conditions~\cite{lopes2021alive2}.
Therefore, \tool complements proof-based checking with execution-backed validation.
\tool introduces \textsc{TestCheck}, a differential validation method over LLVM IR.
Instead of treating execution as random testing, \textsc{TestCheck} uses the strategy $\sigma$ to realize the suspected semantic condition with concrete inputs.
Given $ir_{\text{src}}$ and its transformed version $ir_{\text{tgt}}$, \tool places each IR body into the \verb|@test_func| template in Listing~\ref{lst:difftest-template}, instantiates the placeholder fields \verb|{ret_ty}| and \verb|{args}| according to $\sigma$, and calls the function from \verb|@main| with the same concrete inputs.
Both programs are then executed with LLUBI, a UB-aware LLVM IR interpreter~\cite{online:llubi}.
A mismatch in defined outputs or undefined-behavior is treated as executable semantic evidence.
This design allows \tool to validate cases that are hard for proof-based checking, while still keeping the validation tied to the specific semantic condition proposed \tool.

Algorithm~\ref{alg:guard} shows how \tool deterministically decides whether a validation strategy $\sigma$ can find semantic bugs.
Here, \emph{deterministic} means that \tool cannot report based on textual judgment alone; the guard returns an accepted case $c$ only if $\sigma$ can be materialized into executable evidence.
Otherwise, the algorithm returns $\bot$, and no report is allowed:

\begin{enumerate}[label={\textbf{Step~\arabic*.}}, labelwidth=!, labelindent=30pt, itemsep=5pt, topsep=2pt]
\item \emph{Materialize candidate cases} (line 2):
The guard materializes $\sigma$ into a set of candidate source IR programs $Q$ using PR-related tests $T$ as seeds.
This keeps case generation close to the reviewed patch instead of generating arbitrary LLVM IR from scratch.

\item \emph{Check oracle checkable} (lines 3--11):
For each $ir_{\text{src}} \in Q$, the post-patch compiler $C^{+}$ produces the transformed IR $ir_{\text{tgt}}^{+}$.
The guard first applies \textsc{ProofCheck}; if the result is inconclusive like timeout or \alive errors, it falls back to \textsc{TestCheck}.
Candidates that do not expose differences are discarded.

\item \emph{Check patch triggering} (lines 12--17):
For candidates with a semantic difference, the guard checks whether the difference is introduced by the reviewed patch by comparing $C^{-}$ and $C^{+}$.
Only patch-triggering candidates are accepted as validation cases $c$.
If none is accepted, the guard returns $\bot$.
\end{enumerate}

When the guard accepts a case, \tool reports the validated IR pair $(ir_{\text{src}}, ir_{\text{tgt}}^{+})$ together with concise analyses of the bug trigger and the fix weakness.
The final output is evidence-first rather than a verbose review comment. We provide a concrete example with the guard in Section~\ref{sec:eval-rq4}. 

\begin{table}[tp]
	\centering
	\small
	\caption{Tool taxonomy in \tool.}
	\label{tab:tooling}
	\begin{tabularx}{\columnwidth}{l l X}
		\toprule
		\textbf{Category} & \textbf{Tool} & \textbf{Purpose} \\
		\midrule
		Context
		& \makecell[tl]{\texttt{search}, \texttt{langref}} & Build patch-local context. \\
		\midrule
		Validation
		& \makecell[tl]{\texttt{trans}, \texttt{verify}, \\ \texttt{difftest}} & Turn suspicious strategies into validated evidence. \\
		\midrule
		Management
		& \makecell[tl]{\texttt{workflow}} & Manage review process. \\
		\bottomrule
	\end{tabularx}
\end{table}

\vspace{-2pt}
\section{Implementation}
\label{sec:implementation}

This section discusses two practical implementation aspects when deploying \tool to review LLVM optimization PRs.

\subsection{Dataset and Execution Environment}

\tool uses the GitHub REST API~\cite{online:github-rest} to monitor LLVM PRs with new updates and automatically collect the inputs needed for review.
For each selected PR, \tool extracts the patch diff, changed files, commit metadata, affected optimization pass, and tests attached to or associated with the PR.
It then prepares a patch-specific LLVM environment by building both the pre-patch and post-patch versions of LLVM.
\alive and \llubi are pre-built as validation backends and are invoked by \tool within deterministic validation guards.

\subsection{Compiler-specific Toolkit}

We design a set of compiler-specific tools for \tool to implement the core mechanisms described in Section~\ref{sec:design}, as shown in Table~\ref{tab:tooling}.
Tool calls are one implementation choice, as the same capabilities can also be exposed through other agent workflows, such as agent skills.

To keep the agent's analysis grounded in the reviewed compiler version, \tool restricts the \texttt{search} and \texttt{langref} tools to search the local LLVM source tree and the look up LLVM Language Reference of the current PR.
The agent is not allowed to access the Internet or search  external repositories.
To implement \textsc{ProofCheck} and \textsc{TestCheck} in Section~\ref{sec:design-guard}, \tool provides \texttt{verify} and \texttt{difftest}, which invoke \alive and \llubi under the validation workflow in Algorithm~\ref{alg:guard}.
We also provide \texttt{trans}, which calls the pre-built \opt to perform LLVM IR transformations.
Finally, to reduce uncontrolled behavior in the agentic workflow~\cite{liu2026processcentricanalysisagenticsoftware}, \tool provides a \texttt{workflow} tool to manage review-stage transitions and controlled termination, including exiting with a structured review report.

\begin{table*}[tp]
\centering
\setlength{\tabcolsep}{4pt}
\renewcommand{\arraystretch}{1.1}

\begin{minipage}[t]{0.38\textwidth}
  \vspace{0pt}
  \centering
  \includegraphics[width=\linewidth]{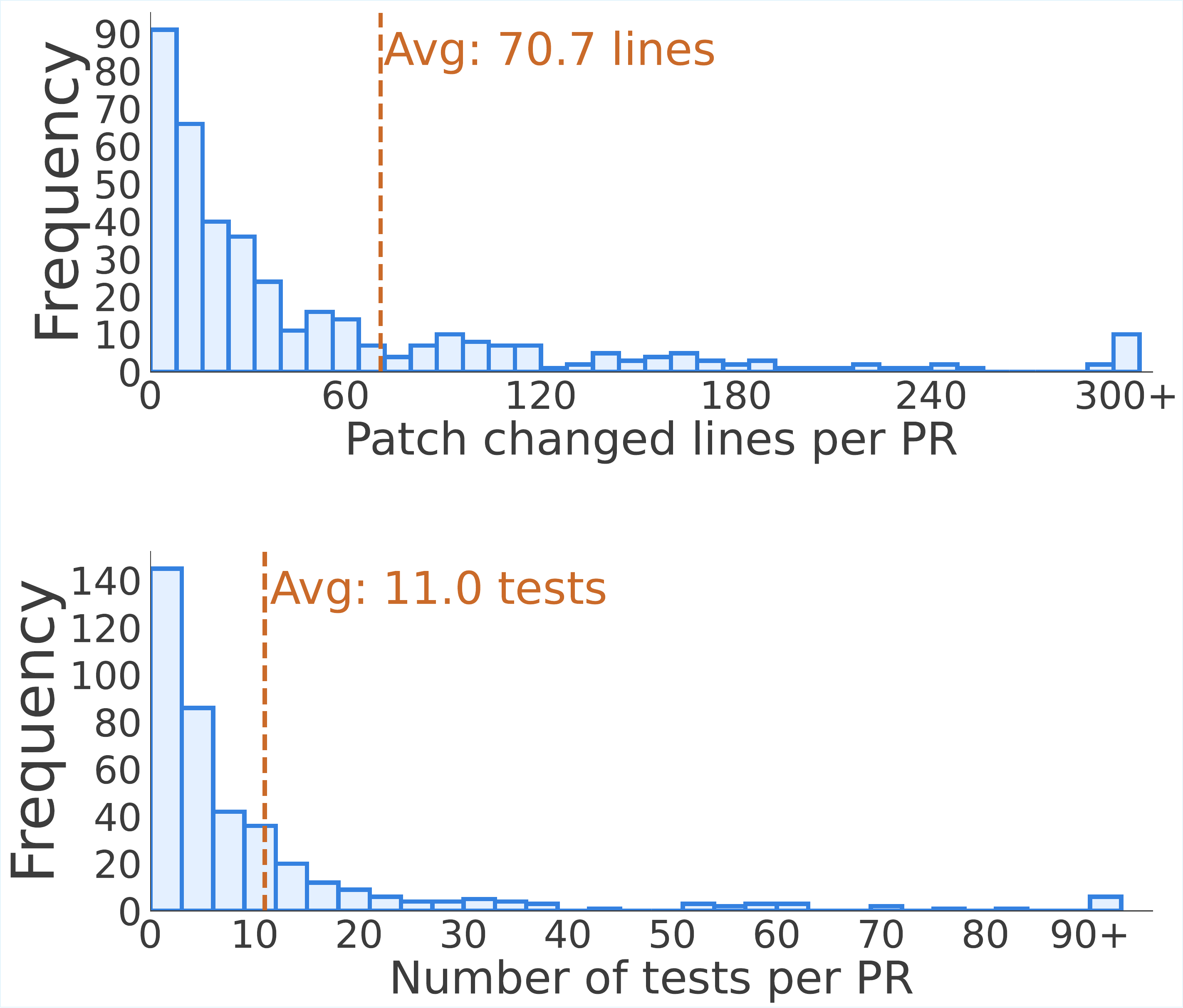}
  \captionof{figure}{Distributions of real-world PRs.}
  \label{fig:dataset}
\end{minipage}
\hfill
\begin{minipage}[t]{0.3\textwidth}
  \vspace{0pt}
  \centering
  \footnotesize
  \captionof{table}{Status of bugs.}
  \vspace{-3pt}
  \label{tab:status}
  \begin{tabular}{c|cc|c}
  \toprule
  \textbf{Status} & \textbf{Open} & \textbf{Closed} & \textbf{Total} \\
  \midrule
  Not Planned & 0 & 3 & 3 \\
  Unconfirmed  & 4  & 0  & 4  \\
  Confirmed & 4  & 7  & 11 \\
  Fixed     & 7  & 26 & 33 \\
  \midrule
  Total     & 15 & 36 & 51 \\
  \bottomrule
  \end{tabular}

  \vspace{1.5em}

  \captionof{table}{Symptoms of bugs.}
  \vspace{3pt}
  \label{tab:symptom}
  \begin{tabular}{c|cc|c}
  \toprule
  \textbf{Symptom} & \textbf{Open} & \textbf{Closed} & \textbf{Total} \\
  \midrule
  Crash           & 3  & 14 & 17 \\
  Miscompilation  & 12 & 22 & 34 \\
  \midrule
  Total           & 15 & 36 & 51 \\
  \bottomrule
  \end{tabular}
\end{minipage}
\hfill
\begin{minipage}[t]{0.3\textwidth}
  \vspace{0pt}
  \centering
  \footnotesize
  \captionof{table}{Affected components.}
  \vspace{0pt}
  \label{tab:affected-llvm-components}
  \setlength{\tabcolsep}{3pt}
  \begin{tabular}{lr}
  \toprule
  \textbf{Component} & \textbf{\#Bugs} \\
  \midrule
  Peephole Optimizations       & 10 \\
  Vectorization Optimization   & 9 \\
  Loop Transformations         & 7 \\
  Value Range Analysis         & 5 \\
  Coroutines                   & 4 \\
  SLP Vectorization            & 3 \\
  Alias Analysis               & 2 \\
  CFG Transformations          & 2 \\
  Inlining                     & 2 \\
  Interprocedural Analysis     & 2 \\
  Interprocedural Optimization & 2 \\
  Constant Propagation         & 1 \\
  Global Value Numbering       & 1 \\
  Pass Management              & 1 \\
  \bottomrule
  \end{tabular}
\end{minipage}
\end{table*}

\section{Evaluation}
\label{sec:evaluation}
In this section, we evaluate the effectiveness and design choices of \tool through the following research questions:

\begin{itemize}[labelwidth=!, labelindent=5pt, itemsep=3pt, topsep=2pt, leftmargin=15pt]
    \item \textbf{RQ1 (Real-world PR review).} \emph{How does \tool perform as a reviewer on the real-world LLVM PRs?}
    \item \textbf{RQ2 (Effectiveness).} \emph{How effective is \tool at identifying semantic bugs on a curated regression benchmark of bisected LLVM PRs compared to different approaches?}  
    \item \textbf{RQ3 (Ablation Analysis).} \emph{How important are obligations and validation guard in \tool, and how do different obligation construction design choices affect its effectiveness?}
    \item \textbf{RQ4 (Case study).} \emph{What bug patterns can \tool uncover, and how does it perform beyond existing tools?}
\end{itemize}

\subsection{Evaluation Setup}

\noindent\textbf{Datasets.} 
We construct two datasets in our evaluation:
\begin{itemize}[labelwidth=!, labelindent=5pt, itemsep=3pt, topsep=2pt, leftmargin=15pt]
    \item \textbf{Real-world dataset.} This dataset consists of all LLVM PRs related to middle-end optimization submitted between December 31st, 2025 and February 28th, 2026, identified by labels, such as \texttt{llvm:transform} and \texttt{llvm:analysis}. In total, it contains 398 PRs, including 70 open PRs and 328 closed PRs. Figure~\ref{fig:dataset} shows the statistics of these PRs, including the number of changed lines in the LLVM source code and the number of tests.
    On average, PRs contain 70.7 lines of code and 11 test cases, indicating a high degree of complexity.
    
    \item \textbf{Regression dataset.} This dataset is built from LLVM middle-end optimization issues labeled as \emph{miscompilation} between January 1st, 2024 and October 1st, 2025. We perform commit-level bisection on each issue to identify the corresponding bug-inducing patch, resulting in 47 cases in this dataset.
\end{itemize}

\smallskip
\noindent\textbf{Obligation Construction.}
We collect LLVM miscompilation bugs between January 2017 and February 2026, extract the associated bug-fixing PRs or commits, and obtain 317 candidate patches in total.
From these patches, \tool constructs validated obligations for 45 optimization passes with 188 validated cases.
For experiments on both the real-world and regression datasets, we explicitly exclude any historical cases that overlap with the evaluation instances before constructing the obligation base, ensuring that no benchmark leakage occurs.

\smallskip
\noindent\textbf{Agent Configuration.}
Each review run is limited to 500 agent rounds and 10M tokens. We allow at most 250 invocations for each tool.
We equip \tool with three different recent models: Gemini-3.1-Pro-Preview-Custom-Tools\cite{online:gemini}, DeepSeek-V3.2~\cite{online:deepseek}, and Qwen3.5-Plus~\cite{online:qwen}.
For mini-SWE-agent~\cite{yang2024swe}, we use its latest V2 version. 
For the closed-source commercial tools GitHub Copilot~\cite{online:copilot}, CodeRabbit~\cite{online:coderabbit} and Greptile~\cite{online:greptile}, we use their Pro versions and keep their default configurations unchanged. For general code agent frameworks, we choose OpenAI Codex~\cite{online:codex} with its latest model ChatGPT-5.5~\cite{online:gpt5.5}.

\smallskip
\noindent\textbf{Environment.} We conducted all our evaluations on one Linux server running Ubuntu 20.04 LTS. It is equipped with an AMD EPYC 7742 64-core CPU and 256GB RAM.

\subsection{RQ1: Real-World PR Review}
\label{sec:eval-rq1}

A central goal of \tool is to support \emph{real-world deployment} by helping relieve the shortage of review capacity in large compiler projects. 
To assess this capability, we deploy \tool in the LLVM community and use it to review 398 pull requests in real-world dataset, including 70 open PRs and 328 closed ones. 
In short, \emph{\tool discovers that more than 21\% of the open PRs and 11\% of the closed PRs in the LLVM compiler are buggy}. Below, we provide a detailed report and analysis.

\para{Number of Bugs.} 
Table~\ref{tab:status} summarizes the status of all bugs reported by \tool. 
In total, \tool reported 51 semantic bugs during the two-month PR review. Among them, 33 bugs (69\%) have already been fixed, and 11 bugs (23\%) have been confirmed by developers but deferred for future resolution. 
These results demonstrate not only the effectiveness of \tool in uncovering semantic bugs, but also the willingness of LLVM developers to take \tool's findings seriously. 
Only three reports are marked as \textit{Not Planned}, showing a low false-positive rate of 6\%.
Two of them are related to a deprecated feature in latest LLVM 23.0 that \tool still considered semantically relevant, while the third stems from an incorrect interpretation of nondeterministic semantics. 
In fact, these cases are acceptable and can be avoided with the continuous development of Agneis, which we will discuss further in Section~\ref{sec:discussion}.
Nevertheless, the significant number of new bugs identified by \tool demonstrates its strong capability.
As noted by a lead maintainer in LLVM: \emph{``Thanks for your work on this! I’ve looked at many of these reports, and I think they’re quite useful. There are few false positives, and the analysis is generally on-point.''}

\para{Symptoms of Bugs.} 
Table~\ref{tab:symptom} summarizes the symptoms of the reported bugs. These bugs fall into two main categories: 
(1) \textit{Crash}: the compiler encounters an internal failure during optimization, such as an assertion violation or other runtime error.
(2) \textit{Miscompilation}: the compiler silently generates incorrect code without any explicit failure. 
As shown in the table, \emph{most of the reported bugs are miscompilation bugs}, which is particularly concerning as miscompilation is widely regarded as the most serious class of compiler bugs~\cite{chen2021compilersurvey}. Detecting such miscompilation bugs requires strong semantic reasoning.

\begin{figure*}[tp]
\centering
\begin{minipage}[t]{0.32\linewidth}
    \centering
    \includegraphics[width=\linewidth]{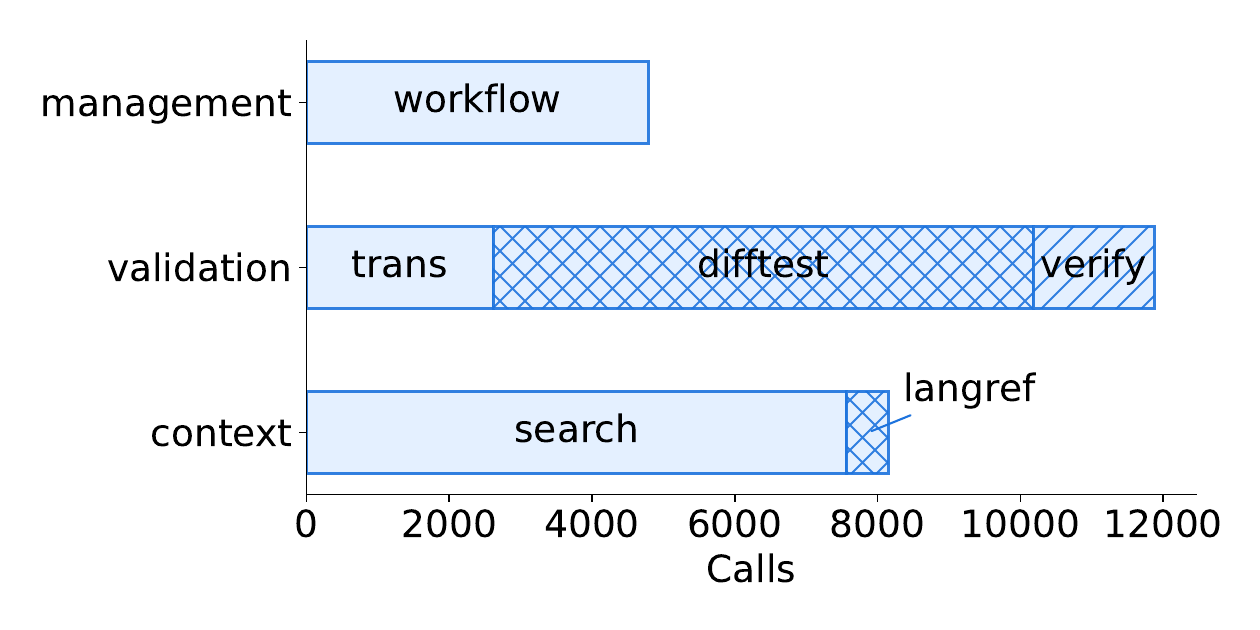}
    \vspace{-8pt}
    \caption{Distributions of tool calls.}
    \label{fig:tool}
\end{minipage}
\hfill
\begin{minipage}[t]{0.32\linewidth}
    \centering
    \includegraphics[width=\linewidth]{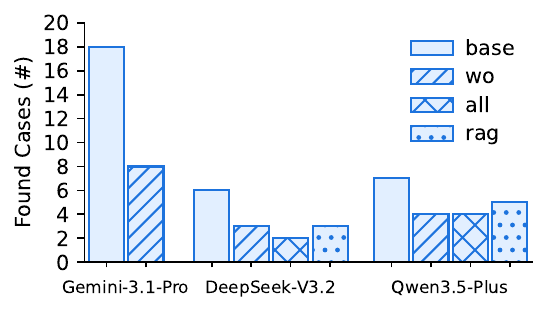}
    \vspace{-8pt}
    \caption{Bugs found with different models.}
    \label{fig:regression-results}
\end{minipage}
\hfill
\begin{minipage}[t]{0.32\linewidth}
    \centering
    \includegraphics[width=\linewidth]{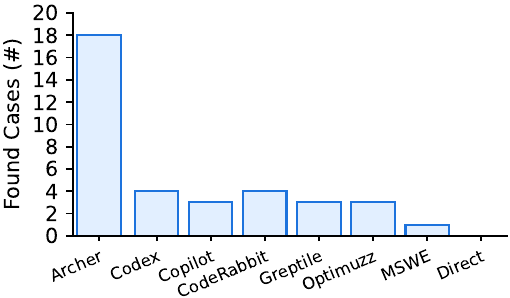}
    \vspace{-8pt}
    \caption{Comparison with different tools.}
    \label{fig:commercial-tools}
\end{minipage}
\end{figure*}

\para{Affected compiler components.} 
Table~\ref{tab:affected-llvm-components} summarizes the compiler components affected by the bugs identified by \tool. 
As shown in the table, \tool uncovers semantic bugs across a diverse set of compiler components, suggesting that its review capability benefits from the pass-level obligation design and extends beyond any single optimization category. 
A substantial portion of the reported bugs arise in peephole optimizations, which is consistent with prior empirical findings on optimization bugs~\cite{zhou2021optimization}. 
We also observe many bugs in loop optimizations and vectorization, where semantic correctness is often harder to validate and many bugs cannot be captured by Alive2 alone, which will be discussed deeper in Section~\ref{sec:eval-rq4}. 

\para{Review process analysis.}
We use Gemini-3.1-Pro from Google as the underlying model, which was among the strongest frontier models available at the time of our study.
Across 398 review cases, the end-to-end review process costs \$988.7 in total, averaging \$2.5 per case.
The review consumes 5,054,201 tokens for each case, and each case takes 877 seconds on average.
This overhead is moderate and remains practical for real-world review settings with bounded budgets.

Figure~\ref{fig:tool} further shows the distribution of tool calls throughout the review process.
As expected, verification tools dominate the overall usage.
Notably, during verification, \texttt{difftest} is invoked more often than \texttt{verify}, highlighting that \textsc{TestCheck} design choice is practical considering the limiation of \textsc{ProofCheck}.
We further analyze the role of context-retrieval tools in successful bug findings.
On average, \tool invokes code search tools to retrieve \(4,800\) additional lines of code for context construction, and consults \texttt{langref} \(1.5\) times to confirm semantic details.

\subsection{RQ2: Effectiveness}
\label{sec:eval-rq2}
In this RQ, we compare \tool with other approaches on the regression dataset that contains PRs with known bugs. This evaluation allows us to quantify \tool's effectiveness.
We configure \tool with different models to understand the impact of the underlying LLM models.  

\para{\tmark Comparing \tool itself using different models.}
We first try to understand how \tool performs when equipped with different LLMs. We provide three alternatives, \ie, one closed-source model Gemini-3.1-Pro, and two open-source models Deepseek-V3.2 and Qwen3.5-plus.

\begin{figure}[tp]
    \centering
    \begin{minipage}[t]{0.49\linewidth}
        \centering
        \includegraphics[width=\linewidth]{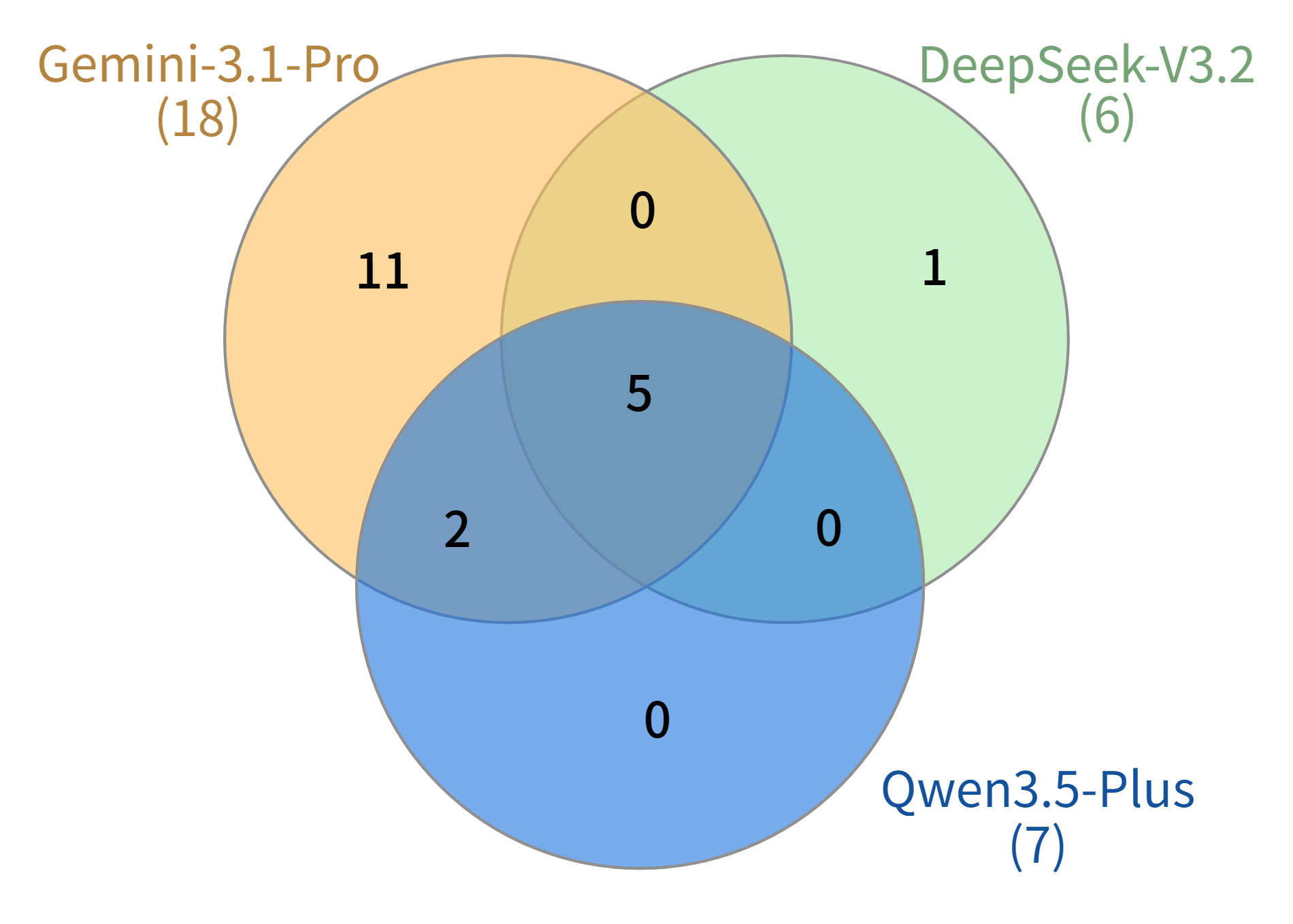}
        \vspace{-5pt}
        \caption{Bug overlap across \tool with different models.}
        \vspace{-10pt}
        \label{fig:unique-results}
    \end{minipage}
    \hfill
    \begin{minipage}[t]{0.49\linewidth}
        \centering
        \includegraphics[width=\linewidth]{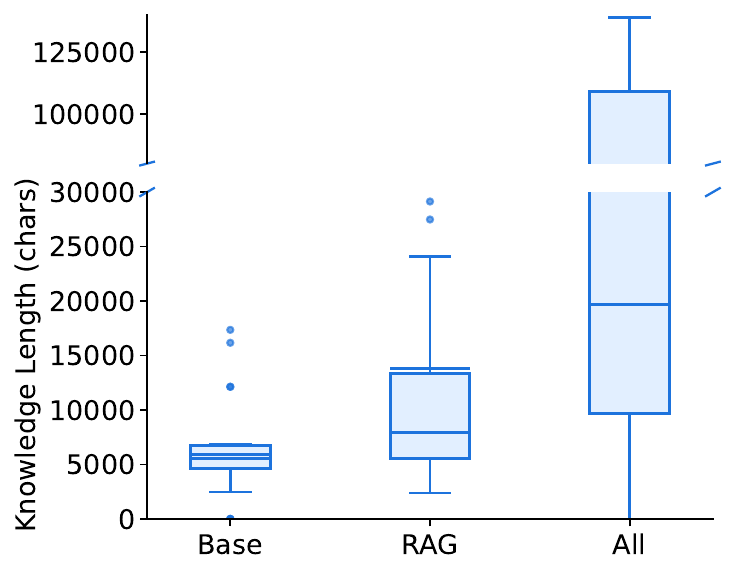}
        \vspace{-5pt}
        \caption{Distributions of different knowledge length.}
        \vspace{-10pt}
        \label{fig:context-length}
    \end{minipage}
\end{figure}

The first three \textit{base} bars in Figure~\ref{fig:regression-results} show that Gemini-3.1-Pro finds the largest number of bugs, followed by Qwen3.5-Plus and DeepSeek-V3.2.
This indicates that stronger foundation models lead to better review outcomes.
For the relatively weaker open-source models, they are also able to find non-trivial numbers of semantic bugs under \tool, suggesting that the effectiveness of \tool is not tied to a single proprietary model.
Figure~\ref{fig:unique-results} shows the bug overlaps. Gemini-3.1-Pro covers all bugs found by the other two models.

\para{\tmark Comparing \tool with general LLM-based methods.}
In this comparison, we want to understand how general LLM and coding agent perform on the compiler review task. We choose two representative baselines as follows.

\begin{itemize}[labelwidth=!, labelindent=5pt, itemsep=3pt, topsep=2pt, leftmargin=15pt]
    \item \textbf{Direct LLM-based query (Denoted as \emph{Direct}).} Deepseek-V3.2 is used directly for review without the agentic framework, obligations, or validation guard.
    \item \textbf{Open-source code agents (Denoted as \emph{MSWE}).} We use mini-SWE-agent with Deepseek-V3.2 serves as the representative open-source agent framework.
\end{itemize}

As shown by the \textit{Direct} and \textit{MSWE} bars in Figure~\ref{fig:regression-results}.
The \textit{Direct} setting finds no bug while the mini-SWE-agent (\textit{MSWE}) setting can only find one bug.
This suggests that although general LLMs and code agents excel at many tasks, it is very hard for them to do complex tasks likes compiler code review, which requires pre-built environment, fine-grained domain obligations and necessary validation suite.

\para{\tmark Comparing \tool with the traditional fuzzing-based tool.}
Optimuzz~\cite{jaeseong2025optimuzz} is a targeted testing tool for LLVM. 
As shown in Figure~\ref{fig:regression-results}, it successfully detects three bugs that are all covered by \tool as well. 
However, Optimuzz fails to start its fuzzing process on 23 out of 47 cases due to the CFG construction failures.
The main reason is that Optimuzz requires to instrument LLVM by matching certain CFG structures, which are often not available in real-world PRs.

\para{\tmark Comparing \tool with the commercial AI review tools.} 
We further compare \tool with commercial agentic review tools on the regression dataset.
These tools are strong general-purpose coding and review systems, and have shown impressive capabilities across many real-world software engineering tasks.
Note that although \tool is currently implemented as a prototype on top of a simple agent framework, its core mechanisms are framework-independent and can be integrated into other agentic review systems.
We evaluate four widely used commercial tools, including Codex, GitHub Copilot, CodeRabbit, and Greptile.
Codex and GitHub Copilot are general-purpose code agents that can be prompted for review, while CodeRabbit and Greptile are designed specifically for code review. 
Since these tools typically produce long natural-language reports rather than executable evidence, we manually inspect each report and check whether it identifies the ground-truth semantic bug.

As shown by the \textit{Codex}, \textit{Copilot}, \textit{CodeRabbit}, and \textit{Greptile} bars in Figure~\ref{fig:regression-results}, all commercial tools perform poorly on the regression dataset.
Most reports focus on superficial implementation concerns, such as missing API parameters, style issues, incomplete comments, or generic maintainability suggestions, while missing the semantic correctness bug introduced by the optimization patch.
A major practical issue is verbosity.
The generated reports often contain many plausible but unverified observations, making manual inspection time-consuming.
After analysis, we find that most issues are false positives or irrelevant to the ground-truth regression.
More seriously, these tools rarely provide executable evidence that links a reported concern to the changed optimization behavior, which makes them impractical for compiler optimization review.

\subsection{RQ3: Ablation Analysis}
\label{sec:eval-rq3}

We conduct the ablation study to examine (1) the contribution of \emph{obligations}, (2) the contribution of \emph{deterministic validation guard}, and (3) the impact of obligation construction design choices. 

\para{\tmark Contribution of obligations.} 
We create a variant, denoted \textit{wo}, by removing all obligations during review.
Figure~\ref{fig:regression-results} shows a consistent gap between the \textit{base} and \textit{wo} settings in all three models, confirming the importance of obligations in semantic compiler review.
Removing obligations reduces the number of bugs found by nearly half across all models.
This result suggests that the model alone is often unable to infer the bug-triggering elements and semantic patterns behind a patch.

\para{\tmark Contribution of deterministic validation guard.} 
The comparison between the \textit{wo} setting of DeepSeek-V3.2 and the \textit{MSWE} bars further highlights the importance of our deterministic validation guard.
Both settings use the same underlying model and framework, but differ in the available tool support.
The \textit{wo} variant of \tool still finds three bugs, whereas \textit{MSWE} finds only one.
This gap indicates that tool access alone is insufficient without carefully designed validation guard.

\para{\tmark Impact of obligation construction design.}
The goal of dynamic obligation construction is to provide concise and effective guidance for both analysis and verification.
Our final design, shown in Algorithm~\ref{alg:pass-knowledge}, uses three LLM stages to transform historical issue-fix instances into structured pass-level obligations.
Theoretically, there can be an arbitrary number of ways of using the historical issue-fix instances to assist code review. 
Below, we provide two straightforward alternatives and compare them with our design.

\begin{itemize}[labelwidth=!, labelindent=5pt, itemsep=3pt, topsep=2pt, leftmargin=15pt]
    \item \textbf{\textit{rag}}: For each PR under review, we use the RAG idea to retrieve historical cases. We use the Jaccard similarity between the patch under review and historical patches to get the most similar top-3 cases, and provide them directly.
    \item \textbf{\textit{all}}: We provide the full set of validated historical cases to the model without any summarization or structuring.
\end{itemize}

The two alternative settings in Figure~\ref{fig:regression-results} correspond directly to the design principles introduced in Section~\ref{sec:design-obligation}.
The \textit{rag} setting represents a static texual alternative that directly retrieves historical issues similar to the patch under review. 
This design weakens the \emph{high-level} and \emph{precise} properties, since the retrieved context remains unverified and tied to concrete past cases rather than abstracting reusable semantic patterns.
The \textit{all} setting provides the full set of validated historical cases without summarization or structuring.
This weakens the \emph{LLM-readable} property, since the model must process a large amount of low-level and case-specific material.
These two alternatives therefore test whether obligations should be abstracted, rather than exposed as raw historical evidence.

As shown in Figure~\ref{fig:regression-results}, the final \textit{base} design consistently outperforms both \textit{rag} and \textit{all} across models.
This indicates that effective review guidance should not simply retrieve similar historical bugs at textual level, but should instead dynamically distill them into compact and reusable semantic obligations.
A further consideration is context efficiency.
Prior work shows that overly long inputs can dilute salient signals even when they remain within the nominal context budget~\cite{dou2026clbenchbenchmarkcontextlearning}.
In agentic review, guidance quality is therefore limited not only by relevance, but also by the practical usability of the context.
Figure~\ref{fig:context-length} compares context length across different settings.
Our constructed obligations, denoted by the \textit{Base} bar, are consistently more compact than both alternatives, indicating a substantial reduction in context size.

\begin{figure*}[tp]
    \centering
    \captionsetup[subfigure]{aboveskip=2pt, belowskip=0pt}

    \begin{subfigure}[t]{.47\textwidth}
        \vspace{0pt}
\begin{lstlisting}[style=irstyle2, xleftmargin=3em]
define i32 @src(i32 %x) {
  loop:
    %iv = phi i32 [ -1, %entry ], ...
    %arith = add i32 %iv, %x
    ; %x=1 -> %arith=0 (non-neg)
    ; But samesign(0, 1) is true (both >= 0)
    %chk = icmp samesign eq i32 %arith, 1
    br i1 %chk, label %exit, label %loop
}
\end{lstlisting}
        \caption{Original: \texttt{samesign(0, 1)} is Valid.}
        \label{fig:rq4-samesign-src}
    \end{subfigure}
    \hfill
    \begin{subfigure}[t]{.51\textwidth}
        \vspace{0pt}
\begin{lstlisting}[style=irstyle2, xleftmargin=5em]
define i32 @tgt(i32 %x) {
    %inv.op = sub i32 1, %x ; x=1 -> inv=0
  loop:
    %iv = phi i32 [ -1, %entry ], ...
    ; LLUBI ERROR: Branch on poison!
    ; -1 is negative, 0 is non-negative.
    %chk = icmp samesign eq i32 %iv, %inv.op
    br i1 %chk, label %exit, label %loop
}
\end{lstlisting}
        \caption{Transformed: \texttt{samesign(-1, 0)} is Poison.}
        \label{fig:rq4-samesign-tgt}
    \end{subfigure}

    \caption{The rewrite \texttt{iv+x==1} $\to$ \texttt{iv==1-x} invalidates \texttt{samesign}. When \texttt{x=1} and \texttt{iv=-1}, the original compare is \texttt{samesign(0, 1)} (valid), but the transformed one is \texttt{samesign(-1, 0)} (poison).}
    \label{fig:rq4-samesign}
    \vspace{-8pt}
\end{figure*}

\subsection{RQ4: Case Study}
\label{sec:eval-rq4}

We present a real-world case study from an LLVM PR in Figure~\ref{fig:rq4-samesign}.
The PR attempts to relax the conditions for hoisting loop-invariant \texttt{add}/\texttt{sub} expressions out of loops when the comparison predicate is \texttt{eq}/\texttt{ne}. While rewriting \texttt{IV + X == C} into \texttt{IV == C - X} is arithmetically sound due to the bijection of modular arithmetic, the patch overlooks the semantics of the \texttt{samesign} flag on the \texttt{icmp} instruction. By keeping the \texttt{samesign} flag unchanged, the transformation introduces a semantic gap.

Guided by obligations for InstCombine pass, \tool generates a targeted strategy focusing on cases related to an \texttt{icmp samesign} instruction. This strategy formulates a specific hypothesis: the rewrite may invalidate the \texttt{samesign} property and thereby produce a \texttt{poison} value. 
\tool then concretizes this strategy into an executable evidence within the deterministic validation guard. \tool successfully synthesizes a concrete loop that exposes the sign mismatch. The final successful evidence, shown in Figure~\ref{fig:rq4-samesign}, fixes the initial loop-carried value \texttt{\%iv = -1} and executes the test with \texttt{\%x = 1}:
\begin{itemize}[labelwidth=!, labelindent=5pt, itemsep=3pt, topsep=2pt, leftmargin=15pt]
    \item \textbf{Original IR (Figure~\ref{fig:rq4-samesign-src}):} The first iteration computes \texttt{\%arith = -1 + 1 = 0}. The comparison \texttt{icmp samesign eq 0, 1} remains valid because both $0$ and $1$ are non-negative, allowing the loop to continue normally after evaluating to \texttt{false}.
    \item \textbf{Transformed IR (Figure~\ref{fig:rq4-samesign-tgt}):} The hoisted invariant becomes \texttt{\%inv.op = 1 - 1 = 0}, and the comparison transforms to \texttt{icmp samesign eq \%iv, \%inv.op}. During the first iteration, this evaluates to \texttt{samesign(-1, 0)}. Since $-1$ (negative) and $0$ (non-negative) have different signs, the \texttt{samesign} constraint is violated. The result becomes \texttt{poison}, triggering immediate undefined behavior.
\end{itemize}

This case also highlights the necessity of \tool's multi-check approach. 
When this loop is submitted to \texttt{verify} tool, it incorrectly reports the transformation as correct with Alive2. 
In contrast, \texttt{difftest} tool using LLUBI clearly identifies the discrepancy, showing that the original program terminates normally while the transformed version fails with a ``Branch on poison'' error. 
This divergence indicates that Alive2's reasoning is imprecise for this specific loop structure. 
Such findings underscore the value of combining \textsc{ProofCheck} alongside \textsc{TestCheck} within \tool.

\vspace{-2pt}
\section{Discussion}
\label{sec:discussion}
\vspace{-2pt}

\para{\tmark Recall gap of \tool.}
Our evaluation on the regression dataset in Section~\ref{sec:eval-rq2} reveals that \tool still has a significant gap in recall, which motivates the following future directions to further improve compiler code review.

\para{\tmark Obligations can be continuously refined.}
Obligations in \tool are not fixed. For failed cases, we can feed the case and the corresponding review output back to the agent and let it revise the relevant obligations automatically. The revised knowledge is then checked against the regression benchmark to verify without hurting performance on existing ones. 

\para{\tmark Review trajectories reveal optimization opportunities.}
Some limitations come from agent behavior rather than missing knowledge. In our traces, the agent sometimes wastes budget on repeated tool calls or low-yield exploration. A natural next step is to train on review trajectories, so that the agent can use its budget more efficiently and behave more like experienced developers during review. 

\section{Related Work}

\para{Targeted Compiler Testing.}
Recent research has proposed a targeted approach for validating and testing specific compiler components. This line of work differs from most prior compiler-testing techniques, including generation-based compiler testing~\cite{yang2011csmith, christopher2015opencl, morisset2013concurrency, livinskii2020yarpgen, livinskii2023yarpgen} and mutation-based testing~\cite{le2014emi, le2015athena, even2023grayc, chen2016classfuzz, christian2012langfuzz, li2024creal}. It leverages semantic information extracted from compiler code to guide testing toward particular components or behaviors.
Optimuzz~\cite{jaeseong2025optimuzz} is the first work to apply directed fuzzing to validate compiler optimizations, combining directed grey-box fuzzing with translation validation.
TargetedFuzz~\cite{zhou2025targetedtestingcompileroptimizations} targets individual optimizations to complement pipeline-based testing. It automatically generates language-agnostic mutators to exercise specific optimization-composition patterns.
MopFuzzer~\cite{xie2025mopfuzzer} focuses on triggering sets of optimizations and implements 13 mutators to maximize optimization interactions in JVM compilers.
\tool differs from targeted compiler testing by focusing on single-patch review, where evidence must be semantically tied to the specific optimization change rather than merely triggering a target transformation.
This difference is also reflected in our evaluation. As shown in Section~\ref{sec:eval-rq2}, targeted testing tools such as Optimuzz are less effective. 

\para{LLMs for Code Review.}
Automated code review has traditionally focused on tasks such as review comment generation and automatic resolution of reviewer comments~\cite{olewicki2024empirical1, wang2025llmasajudge}. 
A recent empirical study~\cite{tufano2024empirical2} emphasizes that current code-review automation techniques are highly task-dependent and that aggregate metrics often obscure where these systems actually succeed or fail. 
With the rise of LLMs, code review systems have increasingly shifted from single-shot generation to more structured workflows. 
CodeAgent~\cite{tang2024codeagent} proposes a multi-agent architecture for code review for tasks such as commit-message consistency checking, vulnerability identification, style validation, and revision suggestion. 
Our setting differs from existing LLM-based code review systems in two key aspects. They generally rely on static patch inspection, which offers limited semantic observability for correctness-critical changes. 
To address these challenges, \tool combines obligations with a deterministic validation guard for executable evidence, which is demonstrated to be useful with results in Section~\ref{sec:eval-rq2}. 

\section{Conclusion}
We presented \tool, a compiler-specific semantic review system for optimization PRs. 
Across 398 recent LLVM PRs, \tool identified 51 semantic bugs, showing that it can improve ongoing review and complement manual inspection by surfacing issues that are easy to miss.
More broadly, we believe this design points to a promising direction beyond compilers. 
For large infrastructure software with complex semantics, effective automated review will likely require not only strong models, but also domain-specific obligations, validation, and workflows that turn semantic suspicion into actionable evidence.

\section*{Data-Availability Statement}
Our research artifacts are publicly available at: \url{https://github.com/cuhk-s3/Archer}. 



\bibliographystyle{IEEEtran}
\bibliography{ref}

\end{document}